\documentstyle[aas2pp4,rotate]{article}

\received{soon}
\accepted{soon after that}
\journalid{}{}
\articleid{}{}

\slugcomment{to appear in A.J.}

\lefthead{Kissler-Patig et al.}
\righthead{Globular Clusters around NGC 1399}

\begin{document}
\input{psfig.sty}

\title{
Towards an Understanding of the Globular Cluster Over--abundance around the
Central Giant Elliptical NGC 1399
}

\author{Markus Kissler-Patig \altaffilmark{1,2}}
\affil{UCO/Lick observatory, University of California,
    Santa Cruz, CA 95064, USA}
\affil{Electronic mails: mkissler@ucolick.org}

\author{Carl J. Grillmair}
\affil{SIRTF Science Center, Mail Stop 100-22, California Institute 
of Technology, Pasadena, CA 91125, USA}
\affil{Electronic mail: carl@ipac.caltech.edu}

\author{Georges Meylan}
\affil{ European Southern Observatory,
        Karl-Schwarzschild-Strasse 2,
        D-85748 Garching bei M\"unchen,
        Germany}
\affil{Electronic mail: gmeylan@eso.org}

\author{Jean P. Brodie}
\affil{UCO/Lick observatory, University of California,
    Santa Cruz, CA 95064, USA}
\affil{Electronic mail: brodie@ucolick.org}

\author{Dante Minniti}
\affil{Lawrence Livermore National Laboratory, Livermore, CA
94550, USA}
\affil{Departamento de Astronom\'\i a y Astrof\'\i sica, P.~Universidad
Cat\'olica, Casilla 104, Santiago 22, Chile}
\affil{Electronic mail: dminniti@llnl.gov}

\author{Paul Goudfrooij \altaffilmark{3}}
\affil{Space Telescope Science Institute, 3700 San Martin
Drive, Baltimore, MD 21218, USA}
\affil{Electronic mail: goudfroo@stsci.edu}

\altaffiltext{1}{Feodor Lynen Fellow of the Alexander von Humboldt Foundation}
\altaffiltext{2}{Current address: ESO, Karl-Schwarzschild-Str.~2, 85748
Garching, Germany. Email: mkissler@eso.org}

\altaffiltext{3}{Affiliated with the Astrophysics Division, Space
Science Department, European Space Agency, ESTEC, Postbus 299, NL-2200
AG Noordwijk, The Netherlands}

\begin{abstract}

We   investigate the kinematics   of   a combined  sample of  74
globular clusters around   NGC 1399. Their high  velocity  dispersion,
increasing  with   radius,   supports their    association  with   the
gravitational potential of the galaxy cluster rather than with that of
NGC 1399 itself.  We find no evidence for rotation in the full sample,
although some indication for rotation in the outer regions. 
The data do not allow us to detect differences 
between the kinematics of the blue and
red  sub--populations of  globular clusters. A  comparison between the
globular cluster systems  of NGC 1399  and  those of NGC  1404 and NGC
1380 indicates that  the globular clusters  in all three galaxies  are
likely  to have formed    via  similar  mechanisms  and  at    similar
epochs. The only property which distinguishes the NGC 1399 globular cluster
system  from these others is that  it is ten  times more abundant.  We
summarize the evidence for associating these excess globulars with the
galaxy cluster rather than with NGC 1399  itself, and suggest that the
over--abundance  can   be explained   by   tidal  stripping, at an early
epoch, of neighboring galaxies and subsequent accumulation  of globulars in  the
gravitational potential of the galaxy cluster.

\end{abstract}



\section{Introduction}

Extragalactic globular clusters have in recent years established
themselves as potential tracers of the formation and evolution of
galaxies (see Ashman \& Zepf 1998 for a recent review). The number of
detailed photometric studies has rapidly increased, and these studies
reveal a number of interesting connections between globular cluster
systems and their host galaxies.  With the recent commissioning of
10m--class telescopes, measuring absorption line indices of a large
number of individual globular clusters has become feasible
(Kissler-Patig et al.~1998; Cohen, Blakeslee, \& Ryzhov 1998).
Colors, magnitudes, total numbers, spatial distributions, radial
density profiles, ages, and metallicities, can now be obtained and
used to constrain the formation history of globular clusters and their
host galaxies.

Another essential source of information for discriminating between
different formation scenarios is the kinematics of globular cluster
systems, as determined from the measured radial velocities of
individual clusters.  For example, the studies of M87 and NGC 1399
found a significantly higher velocity dispersion for the globular
clusters than for the stars (Huchra and Brodie 1987; Mould et
al.~1990; Brodie and Huchra 1991; Grillmair et al.~1994; Cohen \&
Ryzhov 1997; Minniti et al.\ 1998; Kissler-Patig et al.~1998).  Mould
et al.~(1990) demonstrated that, in M87, this was consistent with the
the surface density distribution of globular clusters being more
extended than the surface density distribution of stars.  Grillmair et
al.~(1994) suggested that the globular clusters around NGC 1399 were
reacting to the gravitational potential of the Fornax cluster as a
whole rather than just that of the host galaxy.  In NGC 5128, Harris
et al.~(1988) and Hui et al.~(1995) reported rotation in the globular
cluster system, though only for the metal--rich clusters.  This is
contrary to the findings in NGC 4472 (Sharples et al.~1998) and M87
(Kissler-Patig \& Gebhardt 1998) in which the metal--poor globular
clusters seem to dominate the rotation.

Unfortunately, few  models  exist to compare  with  the (still sparse)
data.   Predictions for  the  kinematic signature in globular  cluster
systems after spiral--spiral mergers have  been presented by Hernquist
\&  Bolte  (1992).  They studied   the kinematics of globular clusters
already present in   the progenitors  and found  that   in the  merger
product  these clusters are expected   to be on   radial orbits in the
inner  regions  and to show    some systemic rotation    far out.  The
kinematics of globular clusters that might have formed during a merger
was not addressed in their study; it would depend on the kinematics of
the in-falling  gas   from which  they  formed.   Other  simulations
(e.g.~Muzzio  1987) studied  the  accretion and  stripping of globular
clusters  in  galaxy  clusters,   but  no  clear  predictions for  the
kinematics of accreted globular clusters were formulated.

The primary  goal  of the  present paper is   to combine  all existing
kinematic data on the  globular cluster system  of NGC 1399  to better
constrain  its  origin. A  secondary  aim  is to  use  the photometric
properties of globular clusters  in  the brightest Fornax  galaxies to
further  constrain   formation  scenarios.    The sample   of   radial
velocities  for globular  clusters around NGC    1399 is compiled  and
presented in Sect.~2.  These are  used  to investigate the   kinematic
properties of the full sample, as well as sub--samples selected on the
basis of  radius and color in Sect.~3.   In  Section 4 we  compare the
kinematics of the  globular clusters with those  of the stars, cluster
galaxies and  X-ray gas,  and derive the  mass-to-light  ratio in  the
outer environs of the galaxy.  In Sect.~5 we compare the properties of
the globular clusters  in NGC 1399 with those  of the globular cluster
systems of the next two brightest  early--type galaxies in Fornax, NGC
1380  and  NGC 1404.  We then discuss   the implications for different
formation scenarios. A  summary and   our  conclusions  are given   in
Sect.~6.


\section{The data}

Our sample is based on the compilations of Grillmair (1992), Grillmair
et  al.\ (1994), Minniti  et  al.\ (1998), and  Kissler-Patig et  al.\
(1998).  Briefly, the data from Grillmair et  al.~were obtained at the
Anglo--Australian      Telescope  with   the Low--Dispersion    Survey
Spectrograph and  the Image Photon  Counting  System in the wavelength
range  3800--4800 \AA\ with $\simeq   13$ \AA\ resolution.  Minniti et
al.~obtained   their data with  the  New  Technology Telescope at  the
European Southern Observatory, using the ESO Multi--Mode Instrument in
the   wavelength  range 6000--9000   \AA\  with   a resolution of  7.5
\AA. Kissler-Patig et   al.~observed with the Low Resolution Imaging
Spectrograph
on the Keck 1 Telescope; their spectra covered a wavelength range from
about 4000 \AA\ to 6100  \AA\ with 5.6 \AA\  resolution. We refer  the
reader to the original  papers for a  more detailed description of the
observations, the data reduction, and the velocity measurements.

Hereafter, we will  refer to the  respective samples as the AAT sample
(Grillmair  1992, Grillmair et  al.~1994), the  NTT sample (Minniti et
al.~1998),   and the Keck   sample (Kissler-Patig  et al.~1998).   The
combined sample  of   74    globular  clusters  is   listed   in
Table~1. For each  globular cluster we  give the ID number (taken from
the  papers with   the prefix  aat/ntt/keck   added respectively), the
equatorial coordinates (B1950), the heliocentric radial velocity (with
the weighted mean when multiple measurements  were available), as well
as the  available  photometric   information.   For  52
globular clusters V$-$I colors  accurate  to 0.035 mag  where obtained
from the  work of Kissler-Patig et  al.~(1997a).  The  B$_j$ magnitude
and  B$_j-$R color were  taken from Grillmair  (1992).  We included in
our combined sample  all objects with  radial  velocities greater than
500 km s$^{-1}$ and less than 2500  km s$^{-1}$ (i.e.~within $3\sigma$
of the mean).   A  list of  positions, velocities  and  colors of  the
74 globular  clusters in electronic form  is available from the
first author. Figure~1 shows the positions with respect to NGC 1399 of
all globular clusters observed;  the  symbols are proportional to  the
difference  between  the globular cluster    velocity and the systemic
velocity of NGC 1399. The two pairs of  larger circles (continuous and
dashed  lines)  show  1 and   5 r$_{\rm  eff}$ for   the stellar light
distribution   of  NGC  1399 (centered)  and  NGC  1404  (in the  SE),
respectively.   In Table~2 we  list the adopted properties of
the three brightest early--type galaxies in  the center of Fornax: NGC
1399, NGC 1404, and NGC 1380.


\section{Kinematic properties of the sample}

\subsection{Velocity distribution of the full sample}

Figure~2 shows a histogram of the globular cluster velocities.  As
already noted by Minniti et al.\ (1998), the velocity distribution
exhibits two peaks roughly centered on the systemic velocity of NGC
1399, although a single Gaussian cannot be rejected at better than the
95\% confidence level according to a $\chi ^2$ or a
Kolmogorov--Smirnov test. The mean velocity of the combined sample of
globular clusters is $1429\pm45$ km s$^{-1}$, similar to the velocity
derived for the stellar component (see Table~2).  No rotation is
detected in the combined sample.  The maximum rotation along any axis
is found to be $74\pm107$ km s$^{-1}$.  A possible cause for the
double peak in the velocity distribution may be contamination of the
sample by globulars belonging to NGC 1404 or NGC 1380.  We discuss
this possibility further below.


\subsection{Spatial distribution of the velocities}

In Figure~3 we plot   the cluster velocities against projected  radius
from the center of NGC 1399. The mean velocity  does not change
significantly with radius (see also Table~3).  However, we note  that
within  4\arcmin\  the velocities  seem  to  cluster  around the  mean
velocity  of the   stellar component  of    NGC 1399, whereas   beyond
5\arcmin\  the globular clusters   seem  to have either  significantly
higher or lower velocities than the  mean.  This is better illustrated
in Fig.~4,    where we have plotted histograms    of the velocities in
concentric  annuli.  Using the {\it  \small ROSTAT}  package (Beers et
al.~1990, Bird \& Beers 1993), we tested the samples for normality and
unimodality.   While   the globular  clusters in  the   inner ring are
consistent  at the 98\% confidence  level  with a normal distribution,
the statistics are inconclusive  in  the middle ring  and inconsistent
with  normality in  the  outer ring   at  the  90\%  confidence level.
However, the number statistics  are rather small, and the distribution
in the outermost annulus is  only  consistent with multi--modality  at
the 80\% confidence level.

It seems reasonable  to   question whether  beyond  6\arcmin\  one  is
starting to sample globular clusters belonging to NGC 1404, which lies
only 9\arcmin\   in projection from NGC  1399.  The  answer comes from
Fig.~1: there is no obvious sign of any concentration of high velocity
clusters in the  direction of NGC 1404.   The second velocity  peak is
therefore probably not due to contamination of  the sample by globular
clusters belonging to NGC 1404.  The  recently derived density profile
for the globular clusters around NGC  1404 (Forbes et al.~1998) indeed
shows that contamination of our sample by  NGC 1404 clusters is likely
to  be  small; half--way between  NGC  1399 and  NGC 1404, the surface
density of globular cluster belonging to NGC 1399 is $\simeq 40$ times
higher than that of NGC  1404. Even at 5  r$_{\rm eff(NGC 1404)}$
($\simeq$ 2.5\arcmin\ ) from
NGC 1404  the contribution of the two  galaxies is  still equal.  This
makes it unlikely for NGC 1404 clusters   to  strongly influence our
sample.
However, we cannot exclude  the possibility that  some of the globular
clusters in our sample  may have been stripped from  NGC 1404 or NGC
1380 at  some point in the past (see Sect.~5.3).


\subsection{Rotation of annular sub--samples} 

We  have searched  for evidence    of   rotation in various    annular
sub--samples.   No significant    rotation   is found   in   the inner
regions. However, in the outer regions ($>$5\arcmin) we find a maximum
rotation  amplitude  of $153\pm93$ km  s$^{-1}$ along a 
position  angle $120\pm40$ degrees (east of  north), roughly along the
isophotal major axis of NGC 1399 ($\simeq  100$ degrees, extrapolating from
the measurements  of   Goudfrooij et  al.\ 1994).  Figure~5  shows the
velocities  of  globular clusters  with  projected radii  greater than
5\arcmin\ versus their position angles. The velocity profile predicted
by the rotation amplitude computed above is shown superimposed.
We ran Monte Carlo simulations (1000 realizations, computing random
data points with the same velocity dispersion, velocity errors and position 
angles as spanned by our sample) to estimate the significance of this result:
the hypothesis of no rotation for this sample is ruled out at the 95\%
confidence level. 

Ostrov  et   al.~(1993), Kissler-Patig  et  al.~(1997a)  and Forbes et
al.~(1998) found   the color distribution    of  NGC 1399   globular
clusters  to exhibit two distinct peaks.  We split the sample into red and 
blue globular clusters (at V$-$I =  1.05, following  Kissler-Patig
et al.~1997a). The small number statistics do not allow us to see any
difference in the kinematics of the blue and red samples. We only note
that our sample is dominated by red clusters inside 5\arcmin\ (mean V$-$I
= $1.16\pm0.02$, where the error is the standard error of the mean), 
while blue and red clusters are about equal in number
beyond 5\arcmin\ from the center of NGC 1399 (mean V$-$I
= $1.04\pm0.04$).

In summary, rotation of the NGC 1399 globular cluster  system as a whole
cannot account for  the possibly double-peaked  nature of the velocity
histogram. However,    we cannot rule    out  relatively high  angular
momentum orbits among  a  subset of the  sample clusters.  Such orbits
would be a natural  consequence of  tidal  interactions or  mergers of
cluster galaxies passing through the  core of the Fornax cluster (cf.\
below).

\subsection{Mean velocities and velocity dispersions of the globular
clusters}

The mean velocities   and velocity  dispersions of globular   clusters
reported in previous  studies (the  AAT,  NTT, and Keck  samples)  are
given in Table~3, together with the values for our combined sample and
the various sub--samples   discussed in  previous sections.  The  mean
velocity  and  velocity dispersions were    computed  using a  maximum
likelihood  dispersion  estimator  (see Pryor   \& Meylan  1993).  The
dispersions of the different samples around a  fixed mean (taken to be
the mean of  the full sample, i.e.,~1429 km  s$^{-1}$) are also given.
We find that the velocity dispersion measured  for the combined sample
agrees  with  those derived individually  for  the AAT,  NTT, and Keck
samples.

More interesting  is the fact that   the velocity dispersion increases
with  radius. This  behavior  is clearly  shown by a locally-weighted,
scatter-plot smoothing fit to  the velocity deviation squared (LOWESS,
see Cleveland \& McGill 1984,  as implemented by Gebhardt et al.~1994)
and shown  as  a solid line   in Fig.~6, on top of   the data for  the
globular clusters. Also   shown in Fig.~6   are the measured  velocity
dispersions   of  the stars,  the   planetary nebulae, and  the Fornax
cluster galaxies.


\section{Dynamics}

\subsection{Are the Stellar and Globular Cluster Measurements Inconsistent?}

The difference between the line--of--sight velocity dispersions of the
globular clusters   and the stars could  be  explained by very
different   density profiles of the   two  components, or by different
degrees of anisotropy.  Wagner et  al.~(1991), Bridges et  al.~(1991),
Kissler-Patig et al.~(1997a) and Forbes  et al.~(1998) found little or
no difference in  the surface density profiles of   the stars and  the
globular clusters, accounting for  $<$10\% of the difference
in the velocity dispersions. A  difference in the degree of anisotropy
is not  excluded and assuming the  stellar  velocity dispersion  to be
isotropic, the higher  velocity  dispersion of the  globular  clusters
would  require that the  tangential   velocity dispersion  of the
globular cluster  system  is at   least  30\% larger  than  its radial
component (following Mould  et   al.~1987).  However, given  that  the
stellar  and globular cluster   data   are sampling different   radial
regimes,   we could equally well  postulate   that  the stars and  the
globular  clusters are   both  isotropic systems.  In  this  case, the
differences in the measured velocity dispersions of  the stars and the
globular clusters would arise from a strong  increase in the potential
field  strength with   radius.  As we   show below,  even assuming the
extreme case  that all  globular  clusters are on circular  orbits, we
find a mass--luminosity ratio several times larger than that which has
been derived for the stellar component within 1.5\arcmin\ of NGC 1399.

\subsection{Can M/L be Constant?}

The spectroscopic measurements of  the integrated stellar light almost
reach  the range in radius where  the innermost  globular clusters are
measured. Together with the data for the planetary nebulae, there is a
consistent  indication of a marked  up--turn in the velocity dispersion
between   1\arcmin\  and 3\arcmin\   from  NGC   1399.   This  is seen
independently in each of the  stellar, planetary nebula, and  globular
cluster data. Regardless  of whether the orbit  shapes are changing or
the Fornax cluster potential is becoming  dominant, this suggests that
all luminous components are similarly affected.  This is not a trivial
result; whereas we find only little 
evidence for rotation in the globular cluster system, Arnaboldi et al.
(1994) claim that the system of  planetary nebulae exhibits a rotation
amplitude of $\sim  250$  km s$^{-1}$ with a PA  $\simeq  -35^\circ$,
i.e.~counter--rotating with respect to the globular clusters. If
real, this might constitute an important clue concerning  the manner in 
which the cD envelope of NGC 1399 was built up.

Following Grillmair  et  al.~(1994), we determine  whether  or not the
increase in  the measured velocity dispersion  at larger radii must be
attributed to an  increase in $M/L$ by  assuming the extreme case that
all of our observed globular clusters are on perfectly circular orbits
(which for  a given potential field strength  will produce the largest
observed   line-of-sight velocity  dispersion).   Assuming a spherical
distribution of mass in hydrostatic  equilibrium, and that the surface
density of  globular clusters and  the stellar  light both follow  the
surface  brightness parameterization of Killeen  \& Bicknell (1988), we
use Equations~(2)  and~(3) of  Grillmair et al.~(1994)  to  solve in a
least-squares sense for a radially--invariant  $M/L$ ratio.   Sampling
the  velocity  dispersion  profile over    the  same distribution   of
projected  radii   as in    our combined  sample,  we   find  that our
observations   are  best   fit    using   $M/L_B    =  50   \pm    15$
M$_\odot$/L$_\odot$,  where   the  quoted  error  reflects   only  the
uncertainty in  velocity dispersion  computed  for the  full sample in
Table~3.

In a way similar to  Minniti et al.  (1998) we  use the projected mass
estimator (Heisler et al.\ 1985), under  the hypothesis of a spherical
distribution  of matter and isotropy  of the velocity dispersion.  For
the   74  globular  clusters, we   obtain   $M_P  =  1.0 \times
10^{13}~M_{\odot}$.  This  mass  estimate, along  with $L  = 4.8\times
10^{10} ~L_{\odot}$ (Grillmair et  al.\  1994, scaled to our   adopted
distance),  gives $M_P/L \sim  208$.   The error  in  this quantity is
dominated  by the  small sample   size and   the different  underlying
theoretical assumptions.   E.g., adopting radial instead  of isotropic
orbits  for the globular   clusters increases the   mass estimate by a
factor of two.

Other  mass  estimators (Heisler et  al.\  1985) applied  to  the same
sample give  the  following results: the   virial  mass $M_{VT} =  8.0
\times 10^{12}~M_{\odot}$ implies $M_{VT}/L \sim 167$, the median mass
$M_M = 6.6 \times 10^{12}~M_{\odot}$ implies $M_M/L \sim 138$, and the
average mass $M_{AV} = 7.8 \times 10^{12}~M_{\odot}$ implies $M/L \sim
162$.   The  virial, projected, and  average  masses  share  the  same
sensitivity to interlopers, i.e.,  globular clusters from the intracluster medium or
from the nearby galaxies.  Eliminating from the sample even a single
such object can decrease the $M/L$ values by as much as 30\% (Minniti
et al. 1998).   These $M/L$ values,  obtained from the globular cluster
system of NGC~1399, are similar  to the corresponding values  obtained
with the same  mass  estimators by Huchra   \& Brodie (1987) from  the
globular cluster system of M87.

Contrary to the velocity dispersion determinations, which suffer only
from the observational errors on the radial velocities, the
mass-to-luminosity estimates accumulate the errors associated with
many various observed parameters, such as the total luminosity within
a given radius, the distance modulus and the radial velocity of the
host galaxy, not to mention the different theoretical assumptions.
Nevertheless, all our $M/L$ determinations (50 -- 200) are larger than
that of a typical old stellar population ($1< M/L <10$), supporting
the existence of a substantial amount of dark matter. We also confirm
that $M/L$ is not constant with radius.

These values are significantly larger than the value $M/L = 17$
determined by Bicknell et al.~(1989), from the stellar velocity
dispersion measurements at radii $<$86\arcsec .  There is
therefore strong evidence for a change in the character of the
gravitational potential at about 2\arcmin\ -- 3\arcmin\ from the
center of NGC 1399.  We conclude, as suggested by Grillmair et
al.~(1994), that most of the globular clusters in our sample are
orbiting in a gravitational potential which is more closely associated
with the galaxy {\it cluster} rather than with NGC 1399 itself.  As
the velocity dispersion profile measured for planetary nebulae at
large radii is very similar to that of the globular clusters, we must
conclude that the stars in the outer cD envelope are similarly
associated with the potential of the Fornax cluster as a whole (see
also Arnaboldi et al.~1996, Mendez et al.~1997).  This would be
consistent with the recent detection of red giant branch stars in the
intergalactic region of the Virgo cluster (Ferguson, Tanvir, \& von
Hippel 1998).

Recent  radial  velocity measurements  by  Cohen \&  Ryzhov  (1997) of
globular clusters  in M87 show similar  velocity dispersions for stars
and globular clusters  within 100\arcsec\ of the center  of M87, but a
steadily rising velocity  dispersion from there outward. These authors
are  similarly driven  to  conclude   that the mass-luminosity   ratio
increases substantially with   radius.   This similarity  may  not  be
surprising as M87    and NGC 1399    both occupy the center   of their
respective cluster potentials.


\subsection{A comparison with the Fornax galaxy cluster}

\subsubsection {The galaxies surrounding NGC 1399}

An  examination  of the velocity  distribution   of 68 Fornax galaxies
studied  by Ferguson (1989)  using the statistical  tests described in
Sect.~3 reveals multiple   peaks of marginal significance   similar to
those apparent    in  Figure~2.  This   motivated    us to   look  for
sub--structure within the Fornax   cluster.  We applied a Dressler  \&
Shectman   (1988) test on  the  entire sample  of  galaxies; with the
exception  of  a  group of  galaxies surrounding  NGC  1316 (Fornax A,
located at  a projected distance of $\simeq$  1.2 Mpc SW of NGC 1399),
we  found no evidence for  sub--structure.  The Fornax cluster appears
to be very homogeneous.  This is in  good accord with the general view
that Fornax is a very compact, relatively relaxed cluster.
Leaving the group  around NGC  1316 out  of the velocity  distribution
considerably  diminishes the  statistical  significance of the  double
peak (to $<50$\%).

The   mean velocity  of  the 57    remaining   galaxies in Fornax   is
$1459\pm40$ km s$^{-1}$, consistent  with the systemic velocity of NGC
1399.  This confirms that NGC  1399 sits at the  center of the cluster
potential well. The  velocity dispersion of  the galaxy  cluster rises
from $\sigma=276$ km s$^{-1}$ in the outer parts (several degrees from
the cluster center) to  $\sigma=413$ km s$^{-1}$ within 24\arcmin\  of
NGC 1399  (den Hartog \& Katgert  1997).  There is very good agreement
(Figure~6) between  the velocity dispersions of   the galaxies and the
globular clusters.

\subsubsection {The X-ray gas}

Jones et al.~(1997) find that the hot X-ray  gas around NGC 1399 has a
temperature of $1.30\pm0.05$ keV (derived  within an annulus extending
from   2' to 18').  This   is energetically  equivalent to a  velocity
dispersion   of $\sigma \simeq   450$ km  s$^{-1}$.  The temperature
profile, converted into a velocity dispersion profile, is shown as the
dashed line in Fig.~6.  Once again  the
agreement  with the globular  cluster measurement is excellent.  Ikebe
et al.~(1996) showed  that the X-ray  gas profile could  be modeled as
the sum of two components, perhaps responding to the potentials of NGC
1399 and the Fornax cluster, respectively. The potential of the galaxy
evidently  falls  off  steeply  and only   clearly  dominates over the
cluster  potential  in the  inner  2\arcmin\   to  3\arcmin  . Beyond
5\arcmin\ or 6\arcmin , the potential of the cluster dominates.


\section{Discussion}

In  order  to better  understand  the origin  of the NGC 1399 globular
cluster  system,   with its  high  velocity  dispersion   and specific
frequency (number of globular clusters  per unit luminosity, Harris \&
van den  Bergh 1981), we pose  the question  whether the properties of
these  globular clusters are peculiar in  any way.  
We  briefly  compare the globular cluster system of NGC 1399 
with those of the next  brightest early--type galaxies in
Fornax, NGC 1380 and NGC 1404. We then discuss our findings within the
framework  of  different  scenarios that   could explain the  globular
cluster over--abundance.

Note that NGC 1399 is traditionally associated with high--$S_N$
galaxies, although Ostrov et al.~(1998) recently suggested that it has
only a moderate over--abundance. The new value is a consequence of 
taking proper account of the light in the extended cD halo. The $S_N$ value 
for NGC 1399, galaxy and cD envelope is still about a factor of two higher 
than for other 
galaxies in Fornax. In particular, the $S_N$ of the cD envelope reaches values 
a factor of 3 higher than the mean of the brightest surrounding
ellipticals, and when the globular clusters are associated with the
``galaxy'' component alone, the $S_N$ reaches values of a factor 5 to 6
higher than the Fornax mean. 

In the following we will distinguish between NGC 1399 including the cD
envelope, the ``galaxy'' component alone (to which we associate the
light of a de Vaucouleurs fit to the central region), and the cD
envelope. $M_V$ and $S_N$ for the different components are listed in Table 2.

\subsection{Photometric and chemical properties of the globular clusters
surrounding NGC 1399}

The globular cluster  system of NGC  1399 has been extensively studied
photometrically (Hanes \& Harris  1986, Geisler \& Forte 1990, Wagner,
Richtler \& Hopp 1991, Bridges, Hanes \&  Harris 1991, Ostrov, Geisler
\&   Forte 1993,  Kissler-Patig  et   al.~1997a, Forbes  et  al.~1998,
Ostrov et al.~1998, Grillmair et al.~1999). The most striking result  of  these
investigations is  the high number of globular clusters surrounding the
galaxy, and the over--abundance  in the number of  globular
clusters per unit light when  compared with the other
``normal'' cluster ellipticals (see Table~2).

Another interesting  finding of  the more  recent  studies is that the
globular    cluster color distribution    has at least two significant
peaks, suggesting   two   or more  sub--populations.   The photometric
studies found that these peaks are separated by about  1 dex in [Fe/H]
and, taking into account the luminosity function, that the majority of
the globular clusters are  as old  (within a  few  Gyr) as the  oldest
globular clusters in the Galaxy.  The nature of these sub--populations was
explored  spectroscopically  by    Kissler-Patig  et al.~(1998).   The
absorption line indices for the two major sub--populations of globular
clusters were   found to be   very  similar to  the  ones observed  in
globular clusters  around M31 and  the  Galaxy, suggesting  similar
formation   epochs  and mechanisms.   Thus  the  majority of  globular
clusters around    NGC  1399 appear   to   be ``normal''  old globular
clusters.  A  small fraction  of   very  red globular   clusters  are
significantly more metal  rich,  and could perhaps  have formed  in  a
later merger.

\subsection{Comparison with the globular cluster systems of NGC 1380 and
NGC 1404}

Do NGC 1380 and  NGC 1404 have globular  cluster systems comparable to
that  of  NGC  1399?  These  two   galaxies are  the   next  brightest
early--type galaxies in Fornax (neglecting currently star--forming NGC
1316). NGC 1380 and NGC 1404 are 1.4 and 1.5 magnitudes fainter,
respectively, than the central cD galaxy, and only 0.3 and 0.4
magnitudes fainter if the light of the cD envelope is neglected (see Table~2). 
These galaxies are projected
only 38\arcmin\ (208 kpc) and 9\arcmin\ (48 kpc) from NGC 1399 and are
presumably well within the embrace of the Fornax cluster potential.

While no spectroscopic observations have yet been made of the globular
clusters surrounding these other two galaxies, a considerable amount of
photometric  work  has   been published  (NGC  1380:  Kissler-Patig et
al.~1997b;  NGC  1404: Hanes  \&  Harris 1986,  Richtler  et al.~1992,
Forbes  et   al.~1998,  Grillmair  et al.~1999).  Qualitatively, the
globular cluster systems of NGC 1399, NGC 1404 and  NGC 1380 look very
similar. All three systems exhibit  broad color distributions with two
(or  possibly three) peaks    (Ostrov  et al.~1993,  Kissler-Patig  et
al.~1997a,b, Forbes  et al.~1998, Grillmair  et al.~1999).
The ratio of  red to  blue globular
clusters is roughly the same  in all three cases,  and between 1 and 2
when integrated  over the  whole  system.  Furthermore,  the two  main
sub--populations peak   at  similar metallicities  (as  derived   from
broad--band colors)   in the three systems.  The  blue populations are
distributed around a  metallicity  typical for the Galaxy  or M31 halo
globular clusters, i.e.,   $-1.5  <$ [Fe/H]  $<  -1.0$   dex. The  red
populations seem to have metallicities  slightly higher than 
disk/bulge globular clusters in the Galaxy (i.e., $-0.5 <$ [Fe/H] $< -0.1$ dex, cf.\
Minniti  1995, although the former  could  be slightly over--estimated from
the broad--band colors, see Kissler-Patig et al.~1998).

Finally, the globular    cluster  luminosity functions  of   all three
systems look similar and  are  well represented  by a Gaussian  with a
$1.2\pm0.2$ mag dispersion and  peak  at about  the same magnitude  to
within the  errors   (see  Table~4).   Assuming that  the   underlying
globular  cluster   mass distributions   are invariant,  this   result
suggests that the globular clusters in the three galaxies have similar
old ages.

The only significant difference between the globular cluster system of
NGC 1399 and those of NGC 1404 and NGC 1380 is  the much higher number
of clusters around NGC   1399.  Quantitatively, NGC 1399   is
surrounded by $5800\pm300$ globular  clusters, while NGC 1404 and  NGC
1380    host only   $800\pm100$   and   $560\pm30$  globular  clusters
respectively (see Table~2).  In  terms of specific frequency, NGC 1399
has $S_N=4.1\pm0.6$ ($S_N=11\pm1$ for the ``galaxy'' component) 
NGC  1404   has $S_N=2.3\pm0.3$, and NGC 1380 has
$S_N=1.5\pm0.2$.  Using the distance in Table~2 and the magnitudes and
colors  of  the galaxies  from the  RC3, the  four smaller early--type
galaxies     in Fornax  NGC  1387,   NGC   1374,  NGC  1379,  NGC 1427
(Kissler-Patig et al.~1997a) have specific frequencies between 2.1 and
3.2 with a  mean  of $2.6\pm0.6$ (where   the error is the  dispersion
around  the  mean). NGC  1399, or rather the center of the Fornax
cluster,  is  outstanding  in its relative
overabundance of globular  clusters, while NGC 1404  and NGC 1380 have
cluster populations close to the mean expected value.

In summary, the globular clusters   around NGC 1399 are  qualitatively
very similar to the ones  surrounding NGC 1380  and NGC 1404, and appear
``normal'' when compared to the globular clusters of the Galaxy. However,
the central galaxy hosts a factor 10 more globular clusters than its
neighbors and has a specific frequency 2 to 3 times higher than expected.
Based on  all the evidence summarized above,  we are led to two conclusions:
{\it i)} the excess globular clusters  do not have unusual properties,
i.e., they are likely to have  formed in the  same kind of process and
at a similar time  as the globular clusters  in NGC 1404 and NGC 1380,
and {\it ii)}  the excess number  of globular clusters is most  likely
linked  to  the special location of   NGC 1399  in   the center of the
cluster.


\subsection{Constraints on scenarios explaining the high specific frequencies 
around central giant ellipticals}

\subsubsection{Current scenarios}

Various scenarios to  explain  the high  $S_N$ values  around  central
galaxies  have   been   comprehensively  discussed   by Blakeslee   et
al.~(1997).   These authors  found   several correlations of  specific
frequency  with  the galaxy     cluster properties,   including  X-ray
temperature, velocity dispersion, and galaxy  distance from the center
of the cluster.  In summary,  {\it  ab initio} scenarios (Harris  1991
and references therein) that associate  the excess numbers of globular
clusters  to  super--efficient globular   cluster formation  in  these
galaxies   fail to explain  the  correlation of   $S_N$, cluster X-ray
properties   and   velocity dispersion. Mergers   as  a  source of the
increase  in $S_N$  (e.g.~Ashman \&  Zepf 1992) also  fail to  explain
these correlations  and have difficulty  explaining the  ratio of blue
and red globular   clusters if the  red  globulars are  thought  to be
responsible  for the high   $S_N$ values (Forbes, Brodie  \& Grillmair
1997).

Scenarios which involve biased formation of  globular clusters in deep
potential wells (West 1993) and intra--cluster globular clusters (West
et  al.~1995) are largely  unconstrained,  and Blakeslee et al.~cannot
rule  them out.  Tidal  stripping  can easily  explain  all   of their
observed correlations; the problem,  as noted by  Blakeslee
et al., is that the stripping simulations  (summarized in Muzzio 1987)
yield too slow an increase  in $S_N$ of the  central galaxy.  Finally,
motivated by the  constant  luminosity of Brightest  Cluster  Galaxies
(BCGs), Blakeslee  et  al.~propose  a new scenario   wherein globular
clusters formed early and their numbers scaled with the available mass
in the galaxy    cluster, whereas  the   luminosity  of the   BCGs  is
relatively independent of the cluster mass and is truncated at a given
luminosity. The  upshot is that  abnormally high $S_N$ values would be
interpreted as a {\it luminosity deficiency\/} rather than a globular
cluster over--abundance (see  also Harris et  al.\ (1998) for  further
support to this idea).

Based on  the observations presented  above,  we concur that excessive
numbers of globular clusters are likely to be linked to the properties
(X-ray temperature; velocity dispersion) of the galaxy cluster and the
special location of high-$S_N$ galaxies.

\subsubsection{Constraints from the properties of the excess globular
clusters} 

Any scenario that purports to  explain the excess of globular clusters
around NGC 1399 must be consistent with the finding that their properties 
are very  similar to those  of globular clusters in NGC 1404 and NGC
1380.     As  discussed  above,   several  sub--populations  have been
identified in  NGC 1399.  The two major  ones, making up  over 90\% of
the system, are old and have mean metallicities of [Fe/H]$\simeq -1.3$
and  $\simeq  -0.6$   dex  respectively.   The ratio    of   these two
populations changes slowly   with   radius, varying from     about 2:1
(red/blue) in the center of  the galaxy to about 1:2  in the cD  halo,
with  an overall average close  to 1:1 (see Kissler-Patig et al.~1997a
and Forbes  et  al.~1998).  In principal, the excess globular
clusters could be made up by metal--rich or metal--poor globular clusters
only. 
Both cases would be inconsistent with NGC 1404 and NGC 1380 having
also metal--poor {\it and} metal--rich clusters, but showing no
over--abundance.
This  suggests that the excess of globular clusters (especially
around the ``galaxy'' component) must be made up of both old, metal--poor 
{\it and\/} old, metal--rich clusters.

This complicates scenarios wherein all the excess clusters are formed
at a very early stage, before the galaxies fully assembled (i.e., West
et  al.~1993;  Blakeslee et  al.~1997;   Harris  et  al.~1998).  Such
scenarios  would require that NGC 1399  have a specific frequency 
about a factor of two  higher than normal, but that all of the  extra
clusters be metal--poor. A mechanism that can explain at least some
increase in the number of red globular clusters seems required.

\subsubsection{Constraints from the properties of BCGs}

Recent   work   by  Arag\'on-Salamanca  et    al.~(1997)  puts further
constraints on  scenarios which require an  excess  number of globular
clusters to have formed at very early times.  These authors see no (or
negative) passive  evolution   in  the  K--band  Hubble    diagram  of
BCGs.   Furthermore, they  detect no young   stellar populations. They
conclude that the total mass of stars in BCGs has grown by a factor of
2 (4) since $z=1$ for $q_0=0$ ($q_0=0.5$),  and cannot be explained by
new stars  forming in mergers  or cooling flows.  BCGs  must therefore
have accreted or annexed mass in the form of old stars (and presumably
old  globular   clusters) since   $z=1$;  either  by  cannibalism  ---
accreting  gas--poor  galaxies ---  or by  stripping of  material from
other galaxies.

\subsubsection{Constraints from the high specific frequency}

Irrespective  of the high {\it total   number\/} of globular clusters,
the high specific frequency, when compared to the neighbor galaxies,
also  needs to be  taken into account.  
If NGC 1399 gained mass simply  by   dissipationless mergers of typical
galaxies surrounding it, then   the specific frequency  would 
{\it decrease} (by  about 30\% for each doubling  of the mass  for NGC
1399), since no neighboring galaxy has a  $S_N$ value much higher than
3. Dissipationless merging therefore seems unlikely to be the main  cause for 
the  growth of the central  galaxy since  $z=1$, unless  the
specific frequency of NGC   1399 was even  higher  at early times.  We
would then need an explanation for the original, even higher, specific
frequency.  It  seems to us more reasonable  to suppose that  the mass
gained  by  NGC   1399  consisted    of accreted  material    with  an
intrinsically high specific frequency of globular clusters that matches
the $S_N\simeq 6$ of the cD halo (see table 2).

Potential  accretion  candidates with high  specific frequencies
include faint dwarf galaxies and galactic halos.  Durrel et al.~(1996)
compiled a table of  $S_N$ values for local  group dwarfs  and several
dwarf ellipticals  in  Virgo.  They find  that  dE galaxies with $M_V<
-15.5$ tend to have a ``normal'' specific frequency ($4.2\pm1.8$), whereas their
two faintest dE,N galaxies have  very high $S_N$ values ($15\pm8$  and
$12\pm9$   respectively).  At least two   other, local group,
low--luminosity dwarf
galaxies are also known to exhibit very high values: Fornax ($29\pm6$)
and  Sagittarius ($25\pm9$).   Miller et   al.~(1998)
recently enlarged the sample   of  dE galaxies with   studied globular
cluster systems  by 24 and  extended  Durrel et al.'s  results in  the
sense  that  their dE,N galaxies   with $M_V<  -15.5$ also have  $S_N$
values around 10 or  above.   However, brighter and  non--nucleated
dEs appear to have more normal specific frequencies. 
Simulations  by  C\^ot\'e et  al.~(1998)  indicate  that the
accretion/stripping of a large number  of dwarf galaxies could explain
the color distributions seen around  giant ellipticals, assuming  that
every  individual   galaxy   forms  globular  clusters    with  a mean
metallicity   proportional to its  size.   However, the  presence of a
large number of such nucleated  dwarf galaxies remnants at the current
epoch  around NGC 1399  can be  ruled   out (Hilker 1998 (chapter  5),
Hilker  et al.~1998a,b and references  therein). Hilker (1998) further
showed in Monte--Carlo simulations that dwarf galaxies (or their  halos) 
cannot explain all the excess
of globular clusters around NGC 1399, even if one assumes an extremely
steep   faint--end of the galaxy  luminosity  function in the past and
efficiently accrete    all the dwarf  galaxies ``missing''   in todays
luminosity function onto NGC 1399.  Therefore,  while accretion of dwarf
galaxies  may have played  a  role  in the  enrichment of the  globular
cluster system of the central galaxy, it cannot be its only cause.

Another  source of  high specific frequency  material  may be galactic
``halos''.  Extended globular cluster   systems  often fall  off  more
slowly   with  radius than   the  stellar  light  profile,  leading to
gradually increasing   specific  frequencies in  the outskirts  of the
distributions.  Recent    studies  have  identified distinct  ``halo''
globular cluster populations in early--type galaxies (Kissler-Patig et
al.~1997b; Lee, Kim \& Geisler 1998) comprised of clusters following a
somewhat shallower density  profile. M87 is  a  clear case,  for which
McLaughlin et   al.~(1994) computed an increase   of $S_N$ from values
around 10  inside  1 r$_{\rm   eff}$ to  25  at  $>$5  r$_{\rm  eff}$.
Unfortunately, such studies   do   not yet exist for   isolated  field
galaxies,   but ``halo''  material  can    be considered a   plausible
candidate   for the  buildup    of  the envelope of NGC 1399 (see   also
Kissler-Patig 1997a and Forbes, Brodie, \& Grillmair 1997). We note in
passing that, in  the galaxy formation  models  of the Searle  \& Zinn
(1978) type,  halos are built  up of fragments  resembling faint dwarf
galaxies.

\subsection{Globular cluster stripping}

Several authors  have  investigated  the importance of   stripping and
harassment  for the formation of     central cD galaxies (see    e.g.,
Dressler 1984  for  an early review).  Following  these  ideas, Harris
(1986)  suggested  that the central  regions of  galaxy clusters could
have  accumulated  globular   clusters  stripped   from their   parent
galaxies. White (1987) suggested that these  clusters could have built
up    the  over--abundant   globular   cluster   systems    around  cD
galaxies. Muzzio (1987) summarized the  results of various simulations
of   globular cluster stripping   in   galaxy  clusters. Some of   his
conclusions   were:  $i$)  stripping of   globular   clusters was most
efficient before the  galaxies and cluster  were virialized, $ii$) the
initial extent of  the halos is  not critical, $iii$) the  location in
the cluster  and the  amount  of dark matter  determine  the number of
encounters and the amount of stripping a galaxy suffers, and $iv$) the
brightest galaxies tend to gain globular clusters, but capture most of
these       from      galaxies     only     somewhat   fainter    than
themselves. Intermediate luminosity  galaxies are the  ones which lose
their globular clusters; faint  or dwarf galaxies are hardly affected.
Unfortunately, the exact  properties of the resulting globular cluster
systems depend too much  on the initial conditions  and the details of
the simulations (which had  rather poor resolution) to allow  detailed
predictions.  We   discuss  the  plausibility  of  the   ``stripping''
scenario in more detail below.

\subsubsection{Do the numbers work out?}

We first determine whether the total number  of globular clusters seen
in  the early--type  galaxies in Fornax   requires that an  additional
formation process be  invoked, or if it  can be explained by  a simple
redistribution  of existing clusters.  In other words, do the galaxies
in the  cluster form a reservoir of  globular clusters large enough to
feed the central galaxy?  To answer  this question, we assume that all
galaxies had the same number of globular  clusters per unit luminosity
to begin with,  and that the excess around  NGC 1399 must be accounted
for  by the  losses suffered by  other  galaxies. What, then,  was the
initial ``universal'' specific frequency in Fornax?

Table~5  summarizes  the result. It   shows  the number  of globular
clusters observed today in each  of brightest early--type galaxies  in
the  center  of the Fornax  cluster, the  number of  globular clusters
initially  resident   in  each galaxy   for   an assumed ``universal''
specific  frequency, and the  number which had  to  be gained or lost.
Assuming an initial $S_N = 3.2$ for NGC 1399, NGC 1404, NGC 1380, NGC 1374, NGC
1379, NGC 1387, and  NGC 1427, one   can account for all  the globular
clusters seen in   these galaxies today.  If we  assume that  the next
seven brightest early--type galaxies (NGC 1336, NGC 1339, NGC 1351, IC
1963, NGC 1380A,  NGC 1381, NGC 1389, taken  from Ferguson 1989),  all
with total magnitudes brighter than $M_V\simeq -19$, suffered the same
losses   as the  other non--central   galaxies,  the  initial specific
frequency  could have been as  low as  3. Note that the
largest losses were suffered by the brightest galaxies.
Including contributions from
dwarf galaxies could lower the  required initial mean value
of $S_N$ still further.

Even if all globular clusters were associated with the ``galaxy''
component, and todays $S_N$ value for NGC 1399 was taken to be $>$10, a similar
calculation would lead to an initial $S_N$ of $\simeq$ 4 for all Fornax
galaxies.

$S_N$ values of 3 to 4 are  not  unusual among early--type,  cluster
ellipticals. The mean for all early--type galaxies with $-19.0 < M_V <
-21.4$  in   various compilations  is  $S_N=3.7\pm3.1$  (Harris 1991),
$S_N=4.1\pm2.9$     (Ashman   \&   Zepf   1998)    and $S_N=4.6\pm2.7$
(Kissler-Patig 1997b), where the  quoted error is the dispersion about
the mean. {\it However}, these numbers need to be confirmed for a sample of 
isolated field  ellipticals  
with  absolute magnitude    $-19<M_V<-22$  before  we can  say    with
confidence that  this  is  likely  to   be representative of   cluster
ellipticals  prior to any  interactions.  Nonetheless, the
point we wish to stress is that {\it the  excess population around the
central galaxy in  Fornax does not require the  production of a  large
number of new globular clusters}.  A simple redistribution, separating
globular clusters from their parent galaxies through tidal encounters,
is sufficient to account  for the observations.   NGC 1399 (or perhaps
more correctly the central potential of the Fornax cluster) would have
gained about 65\% of its current light (assuming for the galaxy without
envelope $M_V=-21.76$, see Table~2), and 75\% of its current
globular clusters through stripping and
harassment.  NGC 1380 and NGC 1404 would have lost 30\% to 50\% of
their globular clusters, and all  other intermediate--size  galaxies  would 
have lost  between  0\% and 30\%   of  their  initial  globular  cluster
populations.

In the above,  we neglected the possible  existence of a population of
intra--cluster globular clusters  (West et  al.~1995, Blakeslee  1996)
which was {\it never} associated with a cluster galaxy. If indeed such
globular clusters formed at early times, then the numbers of globulars
removed from  cluster galaxies needs to  be even less, and the initial
specific frequency  could be even  lower than  the values we have
estimated   above. However, little  is   yet  known concerning such  a
population, making any estimate of its importance uncertain.

Mergers, both gas-rich  and  gas-poor, must  presumably  have occurred
during  the  evolution of  the  Fornax cluster.  However,  rather than
increasing  the specific frequency as   has been often suggested (e.g.,
Zepf \& Ashman 1993), such mergers could just as easily conserve $S_N$
(e.g., van den Bergh 1995).

\subsubsection{Are the other properties of the globular cluster systems
compatible with stripping?}

Kissler-Patig   et al.~(1997a) showed   that  the  combined color
distribution  of  globular  clusters  in  the intermediate  luminosity
galaxies surrounding the central galaxy (NGC 1374, NGC 1379, NGC 1387,
and NGC 1427)  is also consistent with  the observed color distribution  in
NGC 1399, and that this is compatible with the stripping scenario.

Unfortunately, as noted above, few simulations of the stripping of
globular clusters in galaxy clusters exist.  Muzzio (1987) summarized
a number of simulations but was unfortunately not able to make clear
predictions. Globular cluster systems are expected to become extended
in response to the tidal stresses imposed by neighboring galaxies.
However, it is not clear whether by this time we should expect to see
surface density profiles which are tidally truncated, or whether they
should appear extended. 
Tidal  encounters  will naturally lead to  the
growth of  tidal  tails and to   the possible destruction  of numerous
globular clusters (Gnedin \& Ostriker 1997, Combes  et al.~1998).  The
onset of such tidal tails may be detectable in the form of a ``break''
in the surface density profile (Grillmair et al. 1995; 1999).

\subsubsection{A possible scenario}

Currently   available information on the globular  cluster  system of NGC
1399 seems to favor the view that its high specific frequency mainly
results from  tidal stripping of relatively high--$S_N$  material from
neighboring galaxies. If this is the case for other nearby central
cluster galaxies, it
would explain the correlations of $S_N$ with galaxy cluster properties
(Blakeslee et   al.~1997). In  the  case  of NGC 1399,   stripping  is
consistent  with all known properties of  the globular cluster system,
including   the  total number  of  globular   clusters  in the cluster
galaxies.  While this seems sensible,   detailed simulations will   be
needed  to check this suggestion.   A potential remaining problem,  as
already pointed  out by Blakeslee  et al.~1997, is the inefficiency of
stripping  once galaxy clusters are  virialized. This apparent problem
would  however be  resolved if a significant fraction of  the tidal stripping 
occurred during  the  early  phases of  galaxy  formation, when   the  halos of
galaxies  first collapsed and the   old, metal--poor globular  cluster
population formed. However,   tidal  effects would continue  to   pull
clusters  from their host galaxies  until the present day, although at
reduced efficiency.    This,  as  well as  dissipative  post--collapse
merger events, would  allow   for the  addition  of  more  metal--rich
globulars  to the   mix. All  the  constraints discussed  in the above
sections would remain satisfied.


\section{Summary and conclusions}

We have combined data from three separate spectroscopic investigations
to study the kinematics of 74 globular clusters around NGC 1399.
The velocity dispersion   of  the  globular  clusters  increases  with
radius, rising from a value not unlike that  for the outermost stellar
measurements at 2  $r_{\rm eff}$,  to values almost  twice as  high at
$>$5 $r_{\rm eff}$. The  outer velocity dispersion measurements are in
good agreement with the temperature of  the X-ray gas and the velocity
dispersion of galaxies in the  Fornax cluster. We conclude, as already
suggested by Grillmair  et al.~(1994), that  a  large fraction of  the
globular clusters  which we associate with NGC   1399 should rather be
attributed   to  the whole  of  the  Fornax   cluster.  No significant
difference in the  kinematics could be found between  the blue and red
globular cluster sub--populations, but there is some evidence for
rotation in the outer ($>5$\arcmin) regions. 

A  qualitative comparison of the  globular cluster systems of NGC 1399
and neighboring, next--brightest early--type galaxies NGC 1404 and NGC
1380  indicates that  these systems   are indistinguishable from   one
another,  and that there  is no reason  to suppose that they formed at
significantly different epochs or  via a different sequence of events.
The NGC 1399 globular cluster system is distinguished  only in being a
factor of  10 more abundant than the  cluster systems of either of the
other two  galaxies, and having a specific frequency 2 to 3 times higher.
The excess is best understood if a significant fraction of the globular
clusters is indeed associated with the light of the cD envelope.
By association this would mean that the cD envelope 
around NGC 1399 should rather also be associated with the Fornax cluster 
than with the galaxy itself.

We review different scenarios to explain the high specific frequencies
around the central galaxies  and  examine the consequences  of various
existing  constraints for each.  We come  to the conclusion that tidal
stripping of globular clusters  from neighboring galaxies in the early
history of  the  galaxy cluster and the  consequent  buildup of the cD
envelope in  the Fornax  cluster potential  well is the  most likely
explanation.

\acknowledgments

We would like to  thank Ann  Zabludoff for her   improved code of  the
Dressler \&   Shectman test as   well  as useful  comments,  and  Karl
Gebhardt for his LOWESS code.  We are also thankful  to Steve Zepf and
Bill Mathews for interesting  discussions. MKP gratefully acknowledges
the support  of the Alexander von  Humboldt  Foundation. Part  of this
research was funded by the faculty research funds of the University of
California at  Santa  Cruz,  the  HST grant GO.06554.01-95A,   and the
U.S.~Department  of Energy by  Lawrence  Livermore National Laboratory
under Contract W--7405-Eng-48.



\clearpage

\onecolumn

\clearpage

\begin{deluxetable}{l rl r rl rl l}
\tablenum{1}
\tablecaption{Globular clusters around NGC 1399 with measured radial velocities }
\tablehead{
\colhead{ID} & \colhead{RA(1950)} & \colhead{DEC(1950)} & 
\colhead{$v_{\rm helio}$} & \colhead{$V$} & \colhead{$V-I$} & \colhead{$B_j$} & 
\colhead{$B_j-R$} & \colhead{other $v_{\rm helio}$} \\
 & & &  & $\pm 0.02$ & $\pm 0.035$ & $\pm \simeq 0.2$ & $\pm \simeq 0.3$ & 
}
\startdata
aat 1 & 3  35  47.3 &  $-$35  40  21 &  $1121\pm150$ & \nodata & \nodata & 21.7 & 1.31 & \nl
aat 4 & 3  35  49.2 &  $-$35  36  44 &  $2478\pm150$ & \nodata & \nodata & 22.3 & 1.04 & \nl
aat 5 & 3  35  50.1 &  $-$35  38  48 &  $1624\pm150$ & \nodata & \nodata & 21.3 & 1.32 & \nl
aat 6 & 3  35  51.4 &  $-$35  37  52 &  $1186\pm150$ & \nodata & \nodata & 21.9 & 1.13 & \nl
aat 7 & 3  35  52.0 &  $-$35  32  44 &  $1385\pm150$ & \nodata & \nodata & 21.5 & 0.98 & \nl
aat 8 & 3  35  54.8 &  $-$35  38  09 &  $1152\pm150$ & \nodata & \nodata & 22.4 & 0.84 & \nl
aat 10 & 3  35  58.4 &  $-$35  36  03 &  $1068\pm150$ & \nodata & \nodata & 21.9 & 0.91 & \nl
aat 13 & 3  36  03.6 &  $-$35  31  58 &  $1922\pm150$ & 21.15 & 0.97 & 21.9 & 1.05 & \nl
aat 15 & 3  36  05.2 &  $-$35  39  54 &  $1355\pm150$ & \nodata & \nodata & 21.9 & 1.05  & \nl
aat 16 & 3  36  05.9 &  $-$35  34  20 &  $1766\pm150$ & 21.27 & 1.17 & 22.1 & 1.16  & \nl
aat 17 & 3  36  07.8 &  $-$35  34  44 &  $1784\pm150$ & 21.57 & 0.92 & 22.3 & 0.89  & \nl
aat 20 & 3  36  17.8 &  $-$35  38  42 &  $1836\pm150$ & \nodata & \nodata & 21.6 & 1.33  & \nl
aat 21 & 3  36  18.2 &  $-$35  40  52 &  $2085\pm150$ & \nodata & \nodata & 21.9 & 1.27 \nl
aat 25 & 3  36  22.7 &  $-$35  38  12 &  $2182\pm150$ & \nodata & \nodata & 22.4 & 1.19 \nl
aat 26 & 3  36  23.5 &  $-$35  37  25 &  $1646\pm150$ & 20.80 & 1.06 & 21.8 & 1.54 \nl
aat 27 & 3  36  14.7 &  $-$35  33  54 &  $1921\pm150$ & 21.85 & 0.96 & 22.3 & 0.81 \nl
aat 30 & 3  36  24.2 &  $-$35  42  07 &  $1859\pm150$ & \nodata & \nodata & 21.8 & 1.05 \nl
aat 31 & 3  36  24.7 &  $-$35  33  24 &  $1236\pm150$ & 20.59 & 1.01 & 21.4 & 1.23 \nl
aat 33 & 3  36  30.1 &  $-$35  39  10 &  $1350\pm150$ & \nodata & \nodata & 21.6 & 1.48 \nl
aat 34 & 3  36  35.1 &  $-$35  31  15 &  $1701\pm150$ & 20.78 & 1.01 & 21.6 & 1.17 \nl
aat 36 & 3  36  41.3 &  $-$35  41  56 &  $1038\pm150$ & 21.97 & 1.43 & 22.3 & 1.17 \nl
aat 38 & 3  36  43.0 &  $-$35  33  16 &  $ 574\pm150$ & 20.85 & 1.18 & 21.7 & 1.40 \nl
aat 39 & 3  36  44.5 &  $-$35  38  31 &  $1639\pm150$ & 21.19 & 1.27 & 22.0 & 1.32 \nl
aat 40 & 3  36  44.4 &  $-$35  36  49 &  $1539\pm150$ & 20.78 & 1.21 & 21.8 & 1.54 \nl
aat 41 & 3  36  45.7 &  $-$35  38  53 &  $ 571\pm150$ & 20.93 & 1.22 & 21.7 & 1.31 \nl
aat 42 & 3  36  45.8 &  $-$35  37  33 &  $1504\pm150$ & 20.74 & 1.42 & 21.5 & 1.69 \nl
aat 43 & 3  36  46.0 &  $-$35  32  26 &  $1623\pm150$ & 21.23 & 1.13 & 21.9 & 1.17 \nl
aat 48 & 3  36  52.4 &  $-$35  34  33 &  $ 885\pm150$ & 21.71 & 1.16 & 22.4 & 1.19 \nl
aat 49 & 3  36  46.9 &  $-$35  35  44 &  $2026\pm150$ & 21.30 & 1.28 & 22.2 & 1.55 \nl
aat 54 & 3  36  51.9 &  $-$35  33  32 &  $1941\pm150$ & 21.04 & 0.97 & 21.9 & 1.29 \nl
aat 55 & 3  36  54.0 &  $-$35  37  27 &  $1821\pm150$ & 20.94 & 1.00 & 21.6 & 1.01 \nl
aat 56 & 3  36  54.1 &  $-$35  31  25 &  $1206\pm150$ & 21.39 & 1.15 & 21.7 & 1.29 \nl
aat 57 & 3  36  55.4 &  $-$35  31  51 &  $1742\pm150$ & 20.91 & 1.12 & 22.2 & 1.26 \nl
aat 59 & 3  36  59.9 &  $-$35  41  21 &  $1862\pm150$ & 21.12 & 0.75 & 21.5 & 1.24 \nl
aat 62 & 3  37  01.2 &  $-$35  40  49 &  $ 794\pm150$ & 21.25 & 0.89 & 21.2 & 1.35 \nl
aat 66 & 3  37  10.1 &  $-$35  36  36 &  $ 845\pm150$ & 21.17 & 1.12 & 21.8 & 0.98 \nl
aat 67 & 3  37  10.8 &  $-$35  38  42 &  $1343\pm150$ & 21.61 & 0.83 & 22.3 & 1.09 \nl
aat 68 & 3  37  14.3 &  $-$35  37  11 &  $1166\pm150$ & \nodata & \nodata & 21.6 & 1.18 \nl
aat 69 & 3  37  15.1 &  $-$35  33  46 &  $1938\pm150$ & \nodata & \nodata & 22.4 & 0.97 \nl
aat 71 & 3  37  21.5 &  $-$35  39  41 &  $1843\pm150$ & \nodata & \nodata & 22.4 & 1.06 \nl
 & & & & & & & & \nl
ntt 201 & 3 36 44.2 & $-$35 35 40 & $1061\pm135$ &   21.17 & 1.24 &\nodata & \nodata \nl
ntt 203 & 3 37 02.9 & $-$35 34 40 & $ 994\pm073$ &   20.69 & 1.08 &\nodata & \nodata \nl
ntt 208 & 3 36 55.0 & $-$35 31 31 & $1275\pm091$ &   20.91 & 1.12 &\nodata & \nodata \nl
ntt 407 & 3 36 09.8 & $-$35 35 02 & $2107\pm159$ &   20.19 & 1.01 &\nodata & \nodata \nl
ntt 410 & 3 36 17.7 & $-$35 33 45 & $1190\pm094$ &   19.83 & 1.27 &\nodata & \nodata \nl
ntt 414 & 3 36 15.5 & $-$35 32 40 & $1565\pm105$ &   19.56 & 1.09 &\nodata & \nodata \nl
ntt 101 & 3 36 55.4 & $-$35 44 05 & $1270\pm118$ & \nodata & \nodata &\nodata & \nodata \nl
ntt 109 & 3 36 58.9 & $-$35 41 46 & $1426\pm120$ &   21.24 & 1.27
&\nodata & \nodata & $1249\pm103$ , aat 58 $1801\pm150$ \nl
ntt 113 & 3 36 40.7 & $-$35 40 46 & $1440\pm138$ &   21.15 & 1.26 &\nodata & \nodata \nl
ntt 119 & 3 37 01.7 & $-$35 38 18 & $1327\pm121$ &   21.16 & 1.19 &\nodata & \nodata & $1349\pm105$, aat 63 $1282\pm150$ \nl
ntt 122 & 3 37 03.5 & $-$35 37 43 & $1731\pm092$ &   20.77 & 1.06 &\nodata & \nodata \nl
ntt 123 & 3 36 54.1 & $-$35 37 32 & $1307\pm164$ &   20.93 & 1.00 &\nodata & \nodata \nl
ntt 124 & 3 36 43.0 & $-$35 37 14 & $1142\pm189$ &   21.18 & 1.25 &\nodata & \nodata \nl
ntt 125 & 3 36 41.0 & $-$35 37 00 & $1772\pm142$ &   21.02 & 1.17 &\nodata & \nodata \nl
ntt 126 & 3 36 44.7 & $-$35 36 52 & $ 723\pm207$ &   20.76 & 1.22 &\nodata & \nodata \nl
ntt 127 & 3 36 43.5 & $-$35 36 32 & $1811\pm095$ &   21.06 & 1.16 &\nodata & \nodata \nl
 & & & & & & & & \nl
keck 1 & 3 36 13.8& $-$35 39 24.8 & $732\pm032$ & \nodata &\nodata &21.8 &\nodata& \nl
keck 2 & 3 36 14.2& $-$35 38 51.2 & $1094\pm034$ & \nodata &\nodata &22.4 &1.19&\nl
keck 3 & 3 36 09.4& $-$35 37 32.4 & $1571\pm031$ & \nodata &\nodata &22.3 &1.02&\nl
keck 5 & 3 36 13.2& $-$35 37 37.8 & $1775\pm066$ & \nodata &\nodata &21.8 &1.17&\nl
keck 6 & 3 36 17.8& $-$35 37 50.2 & $1386\pm031$ & \nodata &\nodata &22.3 &\nodata&\nl
keck 7 & 3 36 16.7& $-$35 37 01.7 & $1448\pm103$ &21.01 & 1.23 &21.7 &1.31 & $1376\pm84$, aat 28 $1677\pm150$\nl
keck 9 & 3 36 19.2& $-$35 36 28.7 & $1155\pm042$ &21.04 & 1.25 &21.8 &1.33& $1150\pm31$, aat 29 $1280\pm150$\nl
keck 10 & 3 36 21.5& $-$35 36 04.4& $843 \pm045$ &20.55 & 1.05 &21.4 &1.50& $815\pm30$, ntt 406 $917\pm55$, aat 24 $980\pm150$\nl
keck 11 & 3 36 25.0& $-$35 36 28.3& $1338\pm033$ &21.34 & 1.17 &22.2 &1.50& \nl
keck 12 & 3 36 20.3& $-$35 35 15.3& $1734\pm042$ &21.97 & 0.94 &22.4 &1.06& $1736\pm31$, aat 23 $1701\pm150$\nl
keck 13 & 3 36 23.4& $-$35 35 37.2& $1247\pm030$ &21.51 & 0.91 &22.2 & \nodata&\nl
keck 14 & 3 36 24.5& $-$35 35 36.8& $1260\pm066$ &21.17 & 1.37 &22.1 &1.65& \nl
keck 15 & 3 36 23.2& $-$35 34 39.3& $1523\pm030$ &21.26 & 1.04 &22.0 &1.26& \nl
keck 17 & 3 36 26.3& $-$35 34 20.7& $ 866\pm071$ &21.55 & 1.14 &22.4 &1.33& \nl
keck 18 & 3 36 31.4& $-$35 35 05.9& $1688\pm042$ &21.32 & 1.08 &22.3 &1.50& \nl
keck 19 & 3 36 34.5& $-$35 34 52.2& $1150\pm059$ &21.41 & 1.29 &22.3 & \nodata&\nl
keck 20 & 3 36 28.5& $-$35 33 17.0& $1374\pm126$&21.63 & 1.13 &22.3 &0.88& \nl
keck 21 & 3 36 35.7& $-$35 34 24.6& $1154\pm117$&21.15 & 1.09 &22.0 &1.37& $1194\pm98$, aat 35 $1062\pm150$\nl
\enddata
\tablenotetext{}{The AAT data were taken  from Grillmair 1992, the NTT
data from Minniti et al.~1998, and the Keck data from Kissler-Patig et
al.~1998. We used the original ID  numbers, preceded with aat/ntt/keck
respectively. The weighted  mean  velocity was computed when  multiple
measurements  were available,    the original measurements  and  cross
references are given in the last column. $V$ and $V-I$ were taken from
Kissler-Patig  et    al.~1997a,  $B_j$ and   $B_j-R$  were  taken from
Grillmair 1992.}
\end{deluxetable}


\clearpage
\begin{deluxetable}{l c c c l}
\tablenum{2}
\tablecaption{Properties of the galaxies NGC 1399, NGC 1404, and NGC 1380}
\tablehead{
\colhead{} & \colhead{NGC 1399} & \colhead{NGC 1404} & \colhead{NGC
1380} & \colhead{Reference} \\
}
\startdata

RA (1950)        &  03 36 34.0 &  03 36 57.0 & 03 34 31.0   & RC3 \nl
DEC (1950)       & $-$35 36 42 & $-$35 45 18 & $-$35 08 24  & RC3 \nl
l                & 236.71      & 236.95      & 235.93       & RC3 \nl
b                & $-$53.64    & $-53.56$    & $-$54.06     & RC3 \nl
Type (T)         & $-$5.0      & $-$5.0      & $-2.0$       & RC3 \nl
 & & & & \nl
$v_{\rm helio}$ km s$^{-1}$ & $1447\pm12$ & $1929\pm14$ & $1841\pm15$ & RC3 \nl
$(m-M)$    & $31.35\pm0.16$ & $31.35\pm0.16$ & $31.35\pm0.16$ & Della
Valle et al.~1998$^a$ \nl
 & & & & \nl
$V_T$   (mag)    & $8.48\pm0.08^b$ & $10.00\pm0.13$ & $9.93\pm0.10$ &
Ostrov et al.~1998, RC3 \nl
$M_V$ (mag)      & $-22.87\pm0.18$ & $-21.35\pm0.20$ & $-21.42\pm0.19$ &
using $(m-M)$ above \nl
$(B-V)_T$  (mag) & $0.96\pm0.01$ & $0.97\pm0.01$ & $0.94\pm0.01$ & RC3 \nl
$(V-I)_T$ (mag)  & $1.23\pm0.005$ & $1.23\pm0.008$ & $1.18\pm0.004$ & Buta \& Williams 1995 \nl
$R_{\rm eff}$    & $44.7\arcsec$ & $28.5\arcsec$ & \nodata & Goudfrooij et al.~1994 \nl
 & & & & \nl
N$_{\rm GC}$   & $5800\pm300$ & $800\pm100$ & $560\pm30$ & KP97a,F97,R92,KP97b \nl
S$_{\rm N}$    & $4.1\pm0.6$ & $2.3\pm0.3$ & $1.5\pm0.2$ & derived from this table \nl
S$_{\rm N}$(halo) & $6.0\pm1.2$ & \nodata & \nodata & Ostrov et
al.~1998$^b$\nl
S$_{\rm N}$(galaxy) & $11\pm1^b$ & \nodata & \nodata & \nl
\enddata

\tablenotetext{}{RC3:  de     Vaucouleurs   et al.~(1991);    KP97a,b:
Kissler-Patig et al.~1997a,b; F97:  Forbes et al.~1998;  R92: Richtler
et al.~1992}
\tablenotetext{a}{The distance modulus was derived for NGC 1380 and is
assumed to be the same for NGC 1399 and NGC 1404.  It corresponds to a
distance of 18.6 Mpc}
\tablenotetext{b}{Note that for NGC 1399, this includes the cD envelope.
A rough estimate of the ``galaxy'' component can be obtained using the
older value extrapolated from a de Vaucouleurs fit to the inner regions: 
$V_T=9.59\pm0.08$ and $M_V=-21.76\pm0.19$, leading to $S_N=11\pm1$ if
all globular clusters were associated with the galaxy component.}
\tablenotetext{c}{Using our assumed distance modulus, computed for a
annulus with 150\arcsec $<radius<$ 240\arcsec.}

\end{deluxetable}

\clearpage


\begin{deluxetable}{r r l l}
\tablenum{3}
\tablecaption{Mean velocity and velocity dispersion for various samples}
\tablehead{
\colhead{Mean velocity} & \colhead{Standard Deviation} & \colhead{Velocity 
dispersion (mean fixed)} & \colhead{Comment}
}
\startdata
$1518 \pm 91$ & $388\pm54$ & \nodata & AAT sample $^a$\nl
$1353 \pm 79$ & $338\pm56$ & \nodata & NTT sample $^b$\nl
$1293 \pm 71$ & $302\pm51$ & \nodata & Keck sample $^c$\nl
  & & & \nl
$1429 \pm 45$ & $373\pm35$ & \nodata & Full sample \nl
 &  & & \nl
$1484 \pm 128$ & $256\pm86$ & $263\pm92$ &  6 globular clusters within 2\arcmin \nl
$1393 \pm 84$ & $355\pm63$ & $357\pm64$ & 20 globular clusters within 3\arcmin \nl
$1378 \pm 64$ & $371\pm46$ & $375\pm47$ & 41 globular clusters within 5\arcmin \nl
$1498 \pm 68$ & $362\pm52$ & $368\pm54$ & 33 globular clusters outside 5\arcmin \nl
$1421 \pm 83$ & $371\pm63$ & $372\pm63$ & 23 globular clusters outside 6\arcmin \nl
$1515 \pm 134$ & $399\pm101$ & $408\pm107$ & 10 globular clusters outside 8\arcmin \nl
 & &  & \nl
$1611 \pm 87$ & $313\pm69$ & $362\pm104$ & 16 blue globular clusters (V$-$I$<1.05$) \nl
$1322 \pm 58$ & $323\pm45$ & $341\pm51$ & 36 red globular clusters (V$-$I$>1.05$) \nl
$1356 \pm 90$ & $274\pm71$ & $284\pm77$ & 11 red globular clusters outside 5 arcmin \nl
$1643 \pm 154$ & $381\pm117$ & $437\pm174$ & 7 blue globular clusters outside 5 arcmin\nl
\enddata

\tablenotetext{}{
Taken from $^a$ Grillmair et al.~(1994), $^b$ Minniti et al.~(1997), $^c$ 
Kissler-Patig et al.~(1998)
}
\tablenotetext{}{
 $^d$ The dispersion was calculated around the fixed mean of the
full sample (1429 km s$^{-1}$)
}
\end{deluxetable}

\clearpage

\begin{deluxetable}{r c l}
\tablewidth{0pc}
\tablenum{4}
\tablecaption{Peak magnitudes of the globular cluster luminosity function}
\tablehead{
\colhead{Galaxy} & \colhead{$m_V{(\rm peak)}$} & \colhead{Reference}
}
\startdata
NGC 1404  & $24.1 $  $ \pm 0.2 $    &  Richtler et al.~1992 \nl
	  & $23.92$  $ \pm0.20 $    & Blakeslee \& Tonry 1996 \nl
	  & $24.01$  $ \pm0.20  ^a$ & Grillmair et al.~1999 \nl
NGC 1380  & $24.05$  $ \pm 0.25 ^b$ & Blakeslee \& Tonry 1996 \nl
	  & $23.68$  $ \pm 0.11$    & Della Valle et al.~1998 \nl
NGC 1399  & $23.90$  $ \pm 0.09$    & Kohle et al.~1996 \nl
          & $23.83$  $ \pm 0.15$    & Blakeslee \& Tonry 1996 \nl
          & $23.85$  $ \pm 0.30$    & Bridges et al.~1991 \nl
	  & $23.45$  $ \pm 0.16 ^c$ & Geisler \& Forte 1990 \nl
	  & $24.0 $  $ \pm 0.2 $    & Madjesky \& Bender 1990 \nl
	  & $23.73$  $ \pm 0.08 ^a$ & Grillmair et al.~1999 \nl
	  & $23.71$  $ \pm 0.12 ^d$ & Ostrov et al.~1998 \nl
\enddata
\tablenotetext{}{
$^a$ Measured on $V$ data, obtained by transforming $B$ measurements into
$V$ using $(B-I)$ colors.
$^b$ But see comments in Della Valle et al.~1998, 
$^c$ Converted by Geisler \& Forte using $V-T_1=0.45$,
$^d$ Converted from $m_{T_1}$(peak) assuming $(V-T_1)_0 = 0.47$.
}
\end{deluxetable}

\clearpage

\begin{deluxetable}{r r r r}
\tablewidth{0pc}
\tablenum{5}
\tablecaption{Total number of globular clusters in Fornax galaxies}
\tablehead{
\colhead{Galaxy} & \colhead{present N$_{\rm GC}$} & \colhead{initial
N$_{\rm GC}$} & \colhead{\# gained}\\
\colhead{} & \colhead{} & \colhead{} & \colhead{} 
}
\startdata
Initial $S_N=3.2$ & & & \\
NGC 1399 & 5800  &   4527   &  1273 \\
NGC 1380 & 560   &   1195   & --634 \\
NGC 1404 & 800   &   1113   & --313 \\                                   
NGC 1427 & 510   &    510   &     0 \\                                   
NGC 1379 & 310   &    472   & --162 \\                                   
NGC 1387 & 390   &    594   & --204 \\
NGC 1374 & 410   &    410   &     0 \\
total    & 8780   &  8821  & --41 \\
\enddata
\tablenotetext{}{
Assumed  distance to Fornax:  $(m-M)=31.35$. Galaxy luminosities taken
from the RC3 to compute the $S_N$.\\
The present number  of  globular clusters are  taken from  Table 2 and
Kissler-Patig et  al.~1997a. Column 3  shows what  must have been  the
initial number  of globular  clusters for  the  assumed initial $S_N$,
Column 4  shows  the difference  between  the initial number   and the
number of globular clusters presently observed.
}
\end{deluxetable}



\clearpage

\begin{figure}
\psfig{figure=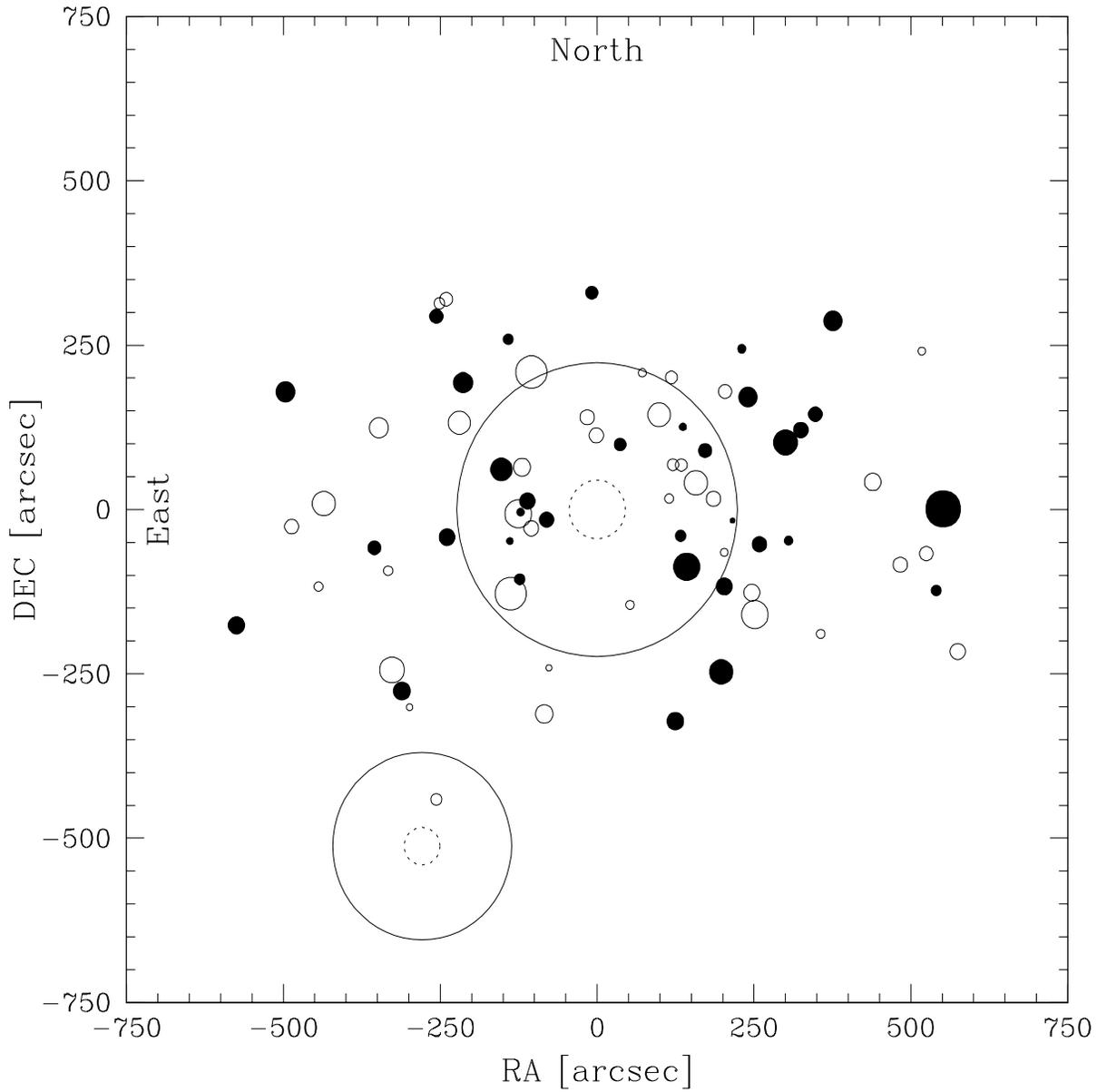,height=16cm,width=16cm
,bbllx=8mm,bblly=57mm,bburx=205mm,bbury=245mm}
\caption{ Positions of the 74 globular  clusters with respect to
NGC 1399 and NGC 1404.  For both galaxies we indicate 1 and 5 r$_{\rm
eff}$.  The  symbols  represent the globular  cluster velocities: open
symbols  are approaching, filled    symbols  receding, with the   size
proportional to the  difference between the globular  cluster velocity
and the mean systemic velocity of NGC 1399.
}
\end {figure}

\clearpage

\begin{figure}
\psfig{figure=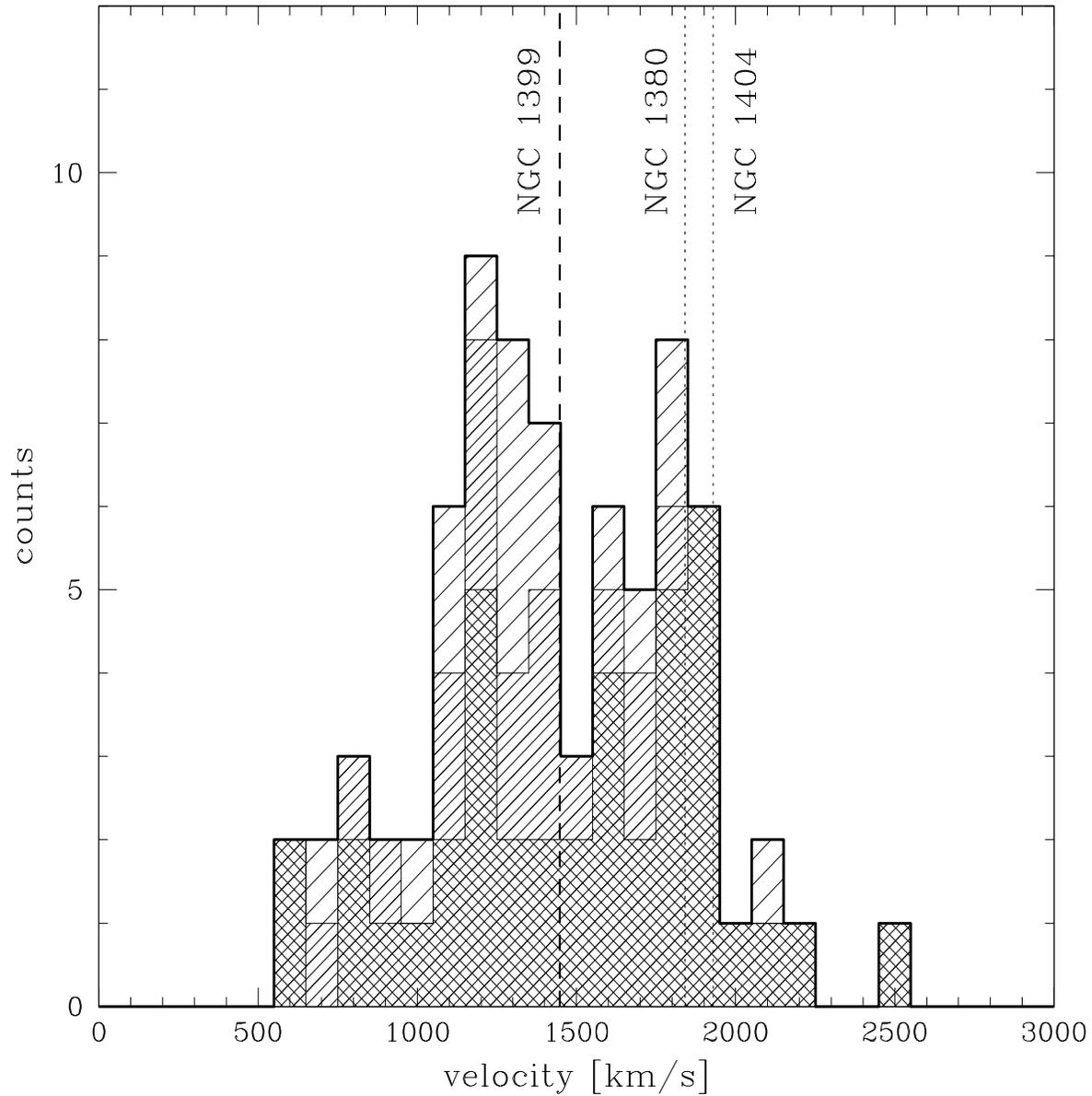,height=16cm,width=16cm
,bbllx=8mm,bblly=57mm,bburx=205mm,bbury=245mm}
\caption{   Histogram    of  74    globular   cluster   velocity
distribution. The long dashed line  indicates the systemic velocity of
NGC 1399, the  short dashed lines the  velocities of NGC  1380 and NGC
1404. The relative contributions of the  AAT sample (crossed regions),
the NTT sample  (narrow hatched regions) and  the Keck sample (hatched
regions) are also shown.
}
\end {figure}

\clearpage


\begin{figure}
\psfig{figure=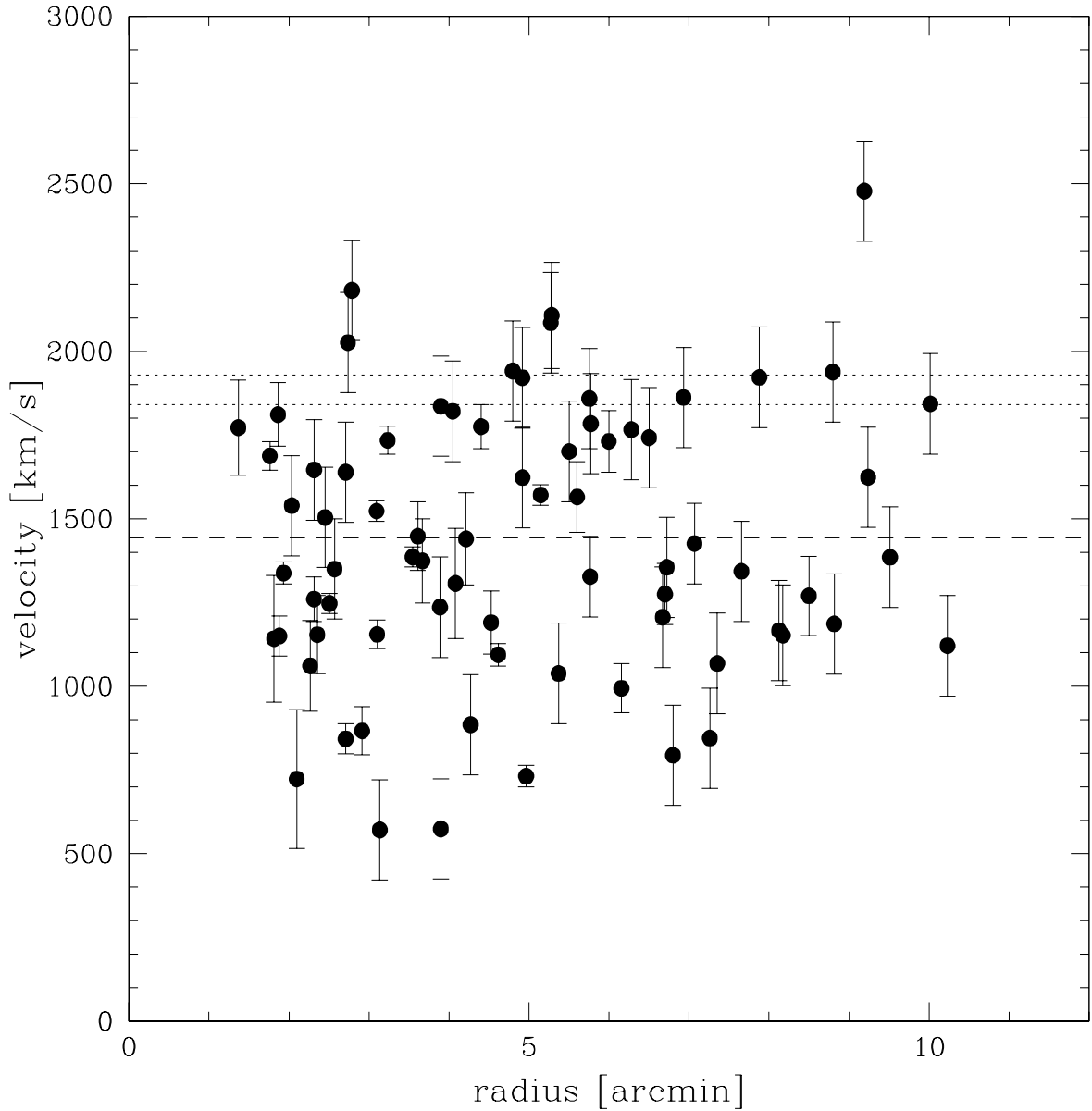,height=16cm,width=16cm
,bbllx=8mm,bblly=57mm,bburx=205mm,bbury=245mm}
\caption{ 
The radial  velocity of  the 74 globular  clusters is  plotted
against their radius from the center of NGC 1399. The long dashed line
indicates the systemic  velocity of NGC 1399,  the short  dashed lines
the velocities of NGC 1380 and NGC 1404.
}
\end {figure}

\clearpage

\begin{figure}
\psfig{figure=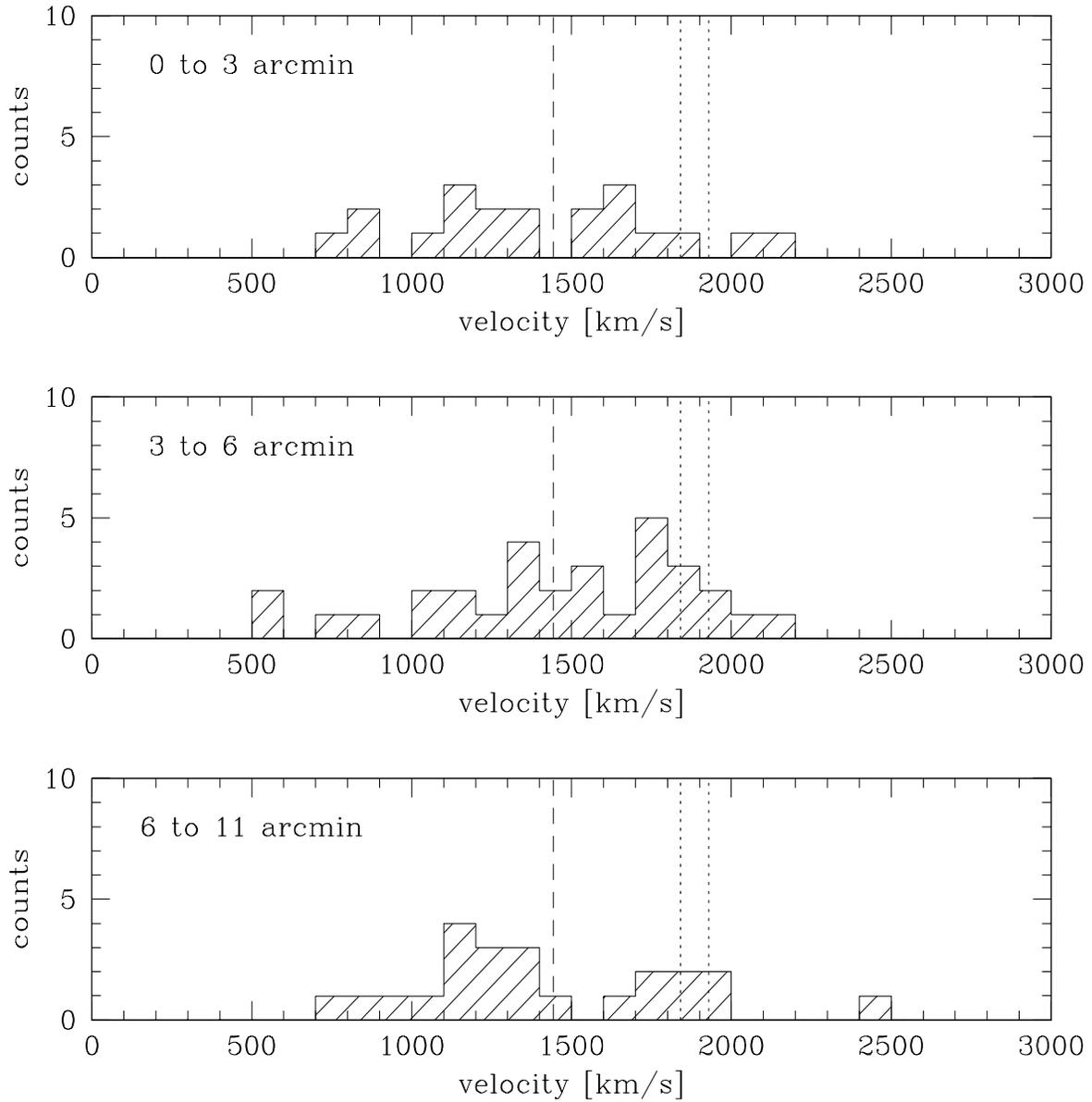,height=16cm,width=16cm
,bbllx=8mm,bblly=57mm,bburx=205mm,bbury=245mm}
\caption{ 
Same as Fig.~2, except  that the full   sample was divided into  three
radial bins: 0'--3' (top), 3'--6' (middle), and 6'--11' (lower panel),
and no distinction was made between the AAT, NTT, and Keck samples.
}
\end {figure}

\clearpage

\begin{figure}
\psfig{figure=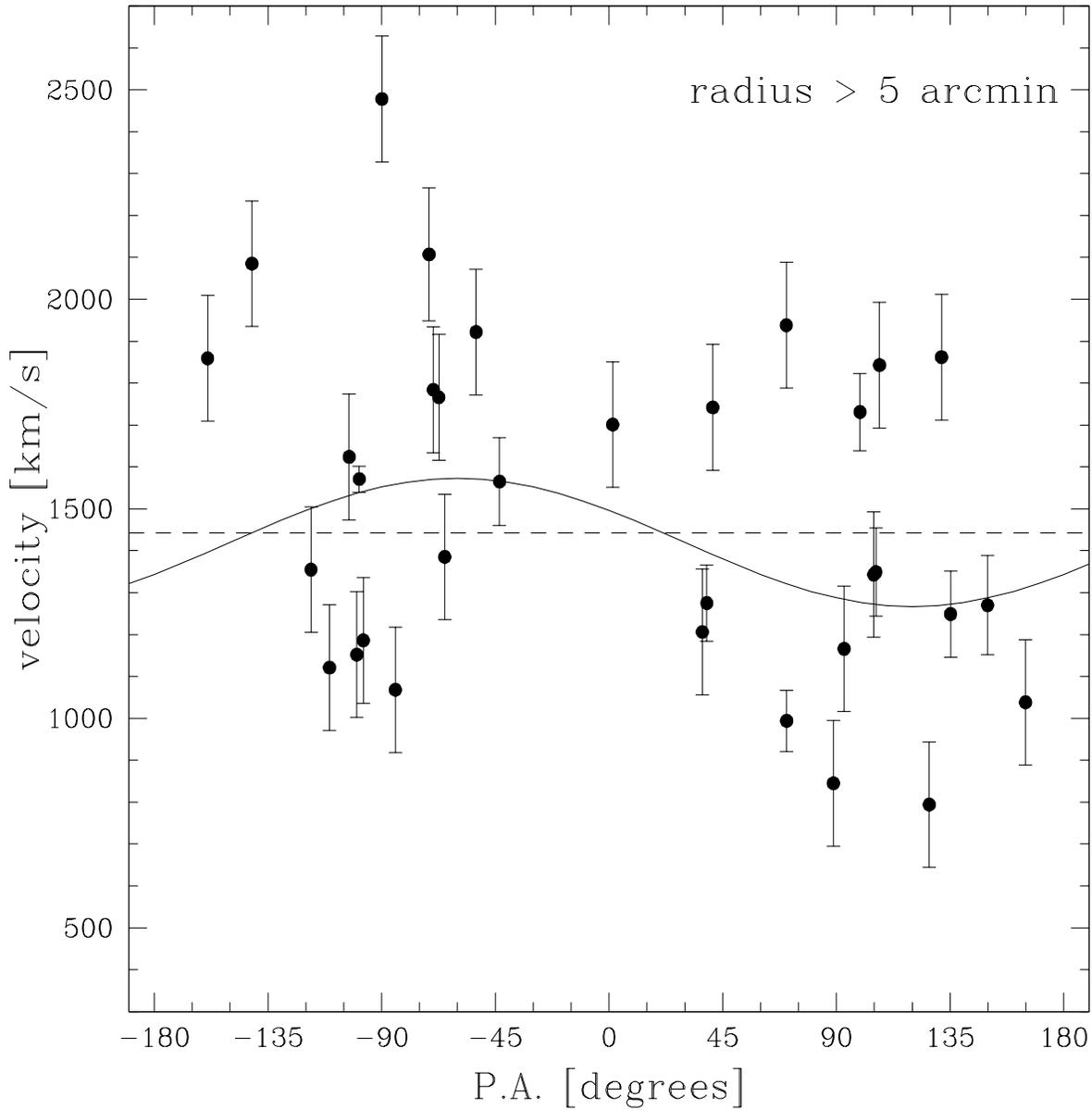,height=16cm,width=16cm
,bbllx=8mm,bblly=57mm,bburx=205mm,bbury=245mm}
\caption{
The velocities of the 33 globular  clusters at a distance greater than
5' from NGC  1399 are plotted against  their position angle. The solid
line  shows the   best  fitted  rotation  (amplitude:   $153\pm93$  km
s$^{-1}$, PA:$120\pm40$ degrees   $\simeq$ the major  axis  of the NGC
1399 isophotes).  The long dashed line  shows the systemic velocity of
NGC 1399.
}
\end {figure}

\clearpage



\begin{figure}
\rotate[r]{
\psfig{figure=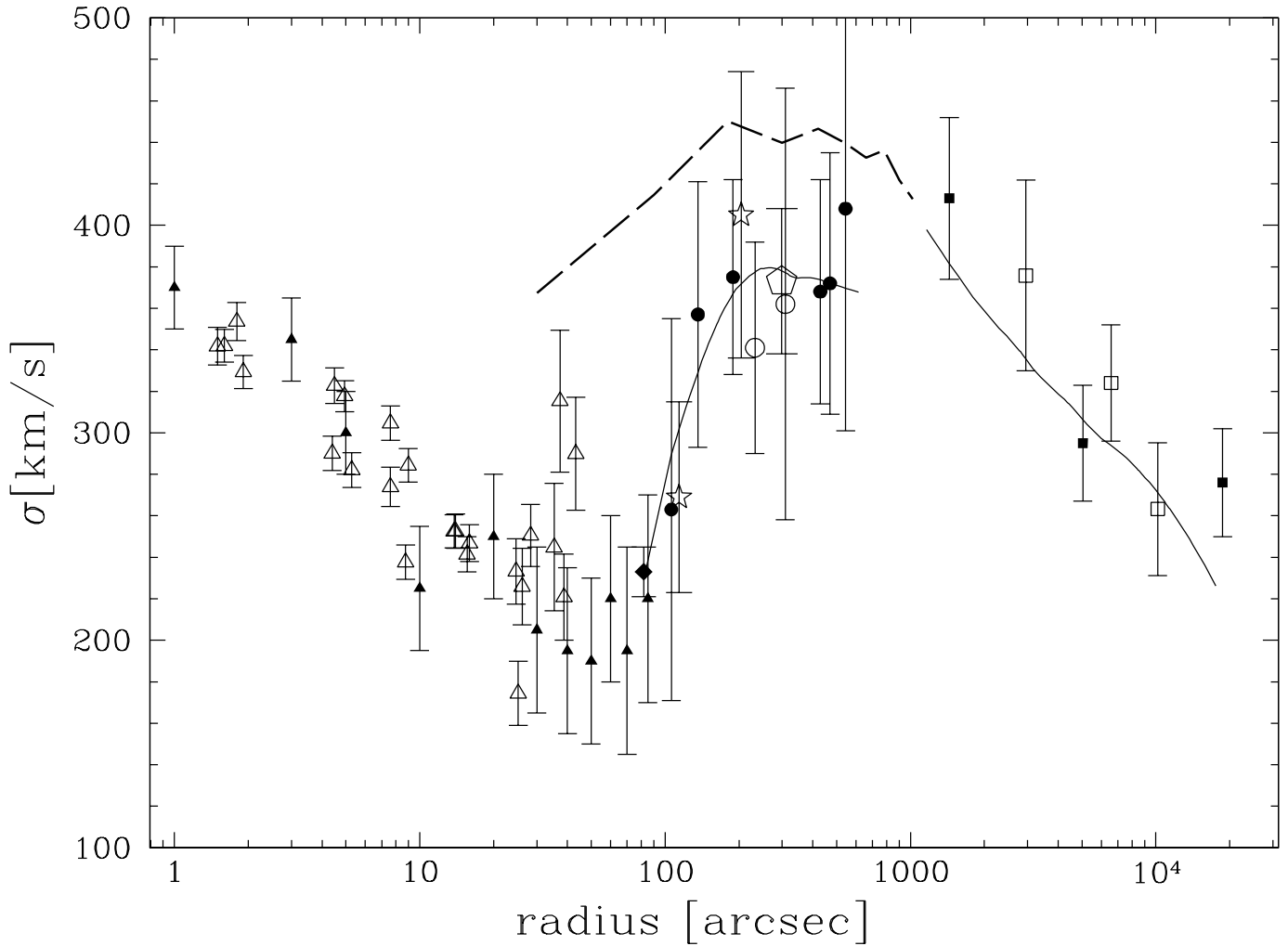,height=14cm,width=18cm
,bbllx=8mm,bblly=8mm,bburx=265mm,bbury=205mm}
}
\vskip -3cm
\caption{ Velocity dispersion versus radius for various
components. Triangles show the velocity dispersion of the stellar
light taken from Franx et al.~(1989, solid symbols), Bicknell et
al.~(1989, open symbols), and Winsall \& Freeman (1993, diamond); 
the stars show the velocity dispersion of
planetary nebulae at two radii taken from Arnaboldi et al.~(1994); the
filled circles mark the velocity dispersion of our different radial
sub--samples, the open circles mark the velocity dispersion of the red
and blue sub--samples, the pentagon marks the velocity
dispersion of the entire sample (data from Table 3); the
squares show the velocity dispersion of Fornax galaxies, taken from
Hartog \& Katgert (1996, filled symbols) , and Ferguson (1989, open
symbols).   The solid line represents the LOWESS fit
to the globular cluster and galaxy data. The dashed line shows the
velocity dispersion profile derived from the temperature of the X-ray
gas (Jones et al.~1997).}
\end {figure}

\end{document}